\documentstyle[amssymb,preprint,aps]{revtex}

\begin{document}
\title{Delayed - Choice Entanglement - Swapping with Vacuum-One Photon Quantum
States.}
\author{Fabio Sciarrino, Egilberto Lombardi, and Francesco De Martini}
\address{Istituto Nazionale di Fisica della Materia, Dipartimento di Fisica, \\
Universita' ''La Sapienza'', Roma, 00185 Italy}
\maketitle

\begin{abstract}
We report the experimental realization of a recently discovered quantum
information protocol by Asher Peres implying an apparent non-local quantum
mechanical retrodiction effect. The demonstration is carried out by applying
a novel quantum optical method by which each singlet entangled state is
physically implemented by a two-dimensional subspace of Fock states of a
mode of the electromagnetic field, specifically the space spanned by the
vacuum and the one photon state, along lines suggested recently by E. Knill 
{\it et al.}, Nature {\bf 409}, 46 (2001) and by M. Duan {\it et al.},
Nature {\bf 414}, 413 (2001). The successful implementation of the new
technique is expected to play an important role in modern quantum
information and communication and in EPR quantum non-locality studies.
\end{abstract}


State entanglement, the most distinctive, fundamental feature of modern
physics is at the heart of the essential non-locality of the quantum world,
i.e. of the irremovable property of nature first discovered in 1935 by
Einstein-Podolsky-Rosen (EPR) and later formally analyzed by J. S. Bell \cite
{1} and recently by L. Hardy \cite{2}. In the context of the modern fields
of quantum information and computation entanglement lies at the core of
several important protocols and methods as for instance the quantum state
teleportation (QST) a fundamental process that has been implemented by
different experimental approaches \cite{3,4}. Very recently quantum
teleportation\ with a unprecedented large ''fidelity''\ has been
experimentally demonstrated by adoption of the new concept of ''entanglement
of one photon with the vacuum'' by which each quantum superposition state,
i.e. ''qubit'', is physically implemented by a two dimensional subspace of
Fock states of a mode of the electromagnetic field, specifically the state
spanned by the QED\ ''vacuum'' and the 1-photon state \cite{5}. This method
requires, as we shall see, an entirely new re-formulation of the Hilbert
space framework supporting the evolution of quantum information, and then
the conception of new devices and methods to implement the transformation
algebra of states and operators. In view of a further clarification of the
new method in the perspective of future more complex applications we
investigate in the present letter the procedure called ''entanglement
swapping'' in which the teleported state itself is entangled, i.e. where the
teleported system does not even enjoy its own state \cite{6}.

Let us first outline the swapping process in the new perspective. It is well
known that the establishment of \ entanglement between two (or more) distant
''quantum systems'' does not necessarily require, as generally believed
after the original EPR approach a direct original interaction between these
ones but it can be realized by merely projecting by an appropriate joint
measurement the independent entangled states pertaining to the separated
systems, \ even in absence of \ any previous mutual interaction. According
to this scenario two separate observers, Alice ($A$) and Bob ($B$)
independently prepare two sets of entangled ''singlets''. They perform on
one ''component'' of each singlet an appropriate test of EPR non-locality,
e.g. a standard Bell-inequality test \cite{1}, a Hardy's ''no-inequality
ladder'' test \cite{2,3} or a ''continuous variables'' homodyne detection
test \cite{7}. The other two components of the singlets are sent to a third
party, Eve ($E$) who performs a joint test at its choice on the components
he received, one from $A$ and one from $B$. By doing that Eve projects (i.e.
''swaps'')\ the state of the two originally non-entangled distant components
in the hands of $A$ and $B$ onto an entangled state. Recently it has been
argued by Asher Peres that the swapping process could be completed by Eve
non necessarily at the time at which the two distant systems were tested by $%
A$ and $B$ but at any retarded ''delayed choice'' time \cite{8}. Indeed,
according to Eve's choice at a later time a fourth ''verification party'',
Victor ($V$) can sort the samples already tested by A and B into subsets and
can verify that each subset behaves as if it consisted of entangled pairs of
distant systems that have never communicated in the past even indirectly via
other systems. This may appear a paradoxical result, as we shall see.

In the present work we report the experimental demonstration of the Peres's
''delayed choice'' process\ by applying the new concept of ''entanglement of
one photon with the vacuum'' \cite{5,9}. Because of \ its novelty let us
outline here the rationale of this approach. The concept of ''non-locality
of a single photon'', first introduced by Albert Einstein in 1927 \cite{10}
has been thoroughly analyzed in the last decade by S. M. Tan et al. \cite{11}%
, by L. Hardy \cite{12} and others in connection with the superposition
state emerging from a beam splitter ($BS$) excited by a single photon at one
of its input ports.\ In our view this state should indeed be interpreted as
an {\it entangled state} by considering that in the domain of optics the 
{\it modes} of the electromagnetic field (e.m.) rather that the photons must
be taken as the ''systems'', or ''components'' to be entangled. This is
consistent with the content of two recent comprehensive theoretical works by
E. Knill {\it et al}. \cite{13} and by M. Duan {\it et al.} \cite{14}. Thus,
any single particle superposition state expressed in the form: $\Sigma _{A}=$
$(2)^{-1/2}\left( \left| 1\right\rangle _{A}\left| 0\right\rangle
_{A^{\prime }}-\text{ }\left| 0\right\rangle _{A}\left| 1\right\rangle
_{A^{\prime }}\right) $ must be interpreted as a ''singlet'' entangling the
mode pair $(k_{A},k_{A^{\prime }})$ which is excited by the Fock states $%
\left| 1\right\rangle $ and $\left| 0\right\rangle $, this last one
expressing the QED vacuum state. If the same states $\left| 0\right\rangle $%
, $\left| 1\right\rangle $ are interpreted respectively as the logic
''zero'', ''one'' information states , the singlet $\Sigma _{A}$ is viewed
as an {\it E-bit}, i.e. an entangled bit of quantum information \cite{15}.
Of course in order to make use of the entanglement present in this picture
we need to use the second quantization procedure of creation and
annihilation of particles and/or use states which are superpositions of
states with different numbers of particles. Another puzzling aspect of this
second quantized picture is the need to define and measure the relative
phase between states with different number of photons, such as the relative
phase between the vacuum and one photon states appearing in Eq.2, below.
That we can associate a relative phase between the {\it vacuum }and anything
else seems most surprising, but it is less so if we recall the more familiar
case of a coherent state, where the relative phase between the different
photon number states in the superposition is reflected physically in the
phase of the classical electric field. To be able to control these relative
phases we need in general, and in analogy with classical computers, to
supply all gates and all sender/receiving stations of any quantum
information network with a common synchronizing {\it clock} signal, e.g.
provided by an ancillary photon or by an ancillary multi-photon, Fourier
transformed coherent state \cite{16}. Optionally in simple cases, as in the
present work, an {\it ad hoc} clock generator is not needed as the mutual
phase information can be retrieved by a linear two mode superposition in a
beam splitter (BS).

An example concerning the present experiment is illustrated by Figure 1
which shows how the non-locality implied by the quantum state of the overall
system could be tested by the distant parties $A$ and $B$ via two coherent
states $\left| \alpha \right\rangle \equiv \left| \left| \alpha \right| \exp
i\theta \right\rangle $ and $\left| \alpha ^{\prime }\right\rangle \equiv
\left| \left| \alpha \right| \exp i\theta ^{\prime }\right\rangle $ that can
operate at the same time as {\it clock states} and as{\it \ local} {\it %
oscillators} (LO) of the corresponding homodyne detectors performing the
same test. The feasibility of\ a similar single photon homodyne technique
has been demonstrated recently \cite{17}.

Figure 1 shows the basic layout of the delayed-choice entanglement swapping
experiment. Pair of photons were generated by Spontaneous Parametric
Down-Conversion (SPDC) excited by a single mode UV CW argon laser in a Type
I $LiIO3$ crystal with the same wavelengths (wl) l = $727,6nm$ and with the
same linear polarizations ($\pi $). Each pair of photons, each of which
associated with an ultrashort optical pulse characterized by a {\it %
coherence time} $\tau _{c}=0,1ps$, was injected into 2 equal 50:50 beam
splitters, $BS_{A}$ and $BS_{B}$ characterized by equal {\it real }%
transmittivity and reflectivity parameters $t=r=2^{-%
{\frac12}%
}$. Precisely, each $BS$ consisted of a 45%
${{}^\circ}$%
$\pi $-rotator followed by a calcite crystal. As it is well known \cite{1,18}%
, the {\it product state}\ character of each pair, $\left| \Phi
\right\rangle =\left| 1\right\rangle _{A}\otimes \left| 1\right\rangle _{B}$
did not imply any inter-particle EPR correlation, in agreement with the data
reported in Figure 2 (open circles). In other words, as far as the dynamics
of the overall system is concerned, each photon pair could have been
supplied equally well by any pair of distant sources. The state $\left| \Phi
\right\rangle $ was transformed by the BS's into the product of two singlets
defined over the pairs of output modes $(k_{A},k_{A^{\prime }})$ and $%
(k_{B},k_{B^{\prime }})$: $\left| \Phi \right\rangle =\Sigma _{A}\otimes
\Sigma _{B}=%
{\frac12}%
\left( \left| 1\right\rangle _{A}\left| 0\right\rangle _{A^{\prime }}-\left|
0\right\rangle _{A}\left| 1\right\rangle _{A^{\prime }}\right) \otimes
\left( \left| 1\right\rangle _{B}\left| 0\right\rangle _{B^{\prime }}-\left|
0\right\rangle _{B}\left| 1\right\rangle _{B^{\prime }}\right) $. The pure
state $\left| \Phi \right\rangle $ may be expressed as a sum of products of
Bell states defined in the two 2-dimensional Hilbert subspaces spanned by
the state eigenvectors to be measured respectively by the couple (Alice,
Bob) and by Eve: 
\begin{equation}
\left| \Phi \right\rangle =\Sigma _{A}\otimes \Sigma _{B}=%
{\frac12}%
\left[ \Phi ^{+}\otimes \Phi _{E}^{+}-\Phi ^{-}\otimes \Phi _{E}^{-}-\Psi
^{+}\otimes \Psi _{E}^{+}+\Psi ^{-}\otimes \Psi _{E}^{-}\right]
\end{equation}
and the Bell states defined in the corresponding 2-d Hilbert sub-spaces are 
\cite{5}: 
\begin{eqnarray}
\Phi ^{\pm } &=&\frac{1}{\sqrt{2}}\left( \left| 0\right\rangle _{A}\left|
0\right\rangle _{B}\pm \left| 1\right\rangle _{A}\left| 1\right\rangle
_{B}\right) ,\Psi ^{\pm }=\frac{1}{\sqrt{2}}\left( \left| 0\right\rangle
_{A}\left| 1\right\rangle _{B}\pm \left| 1\right\rangle _{A}\left|
0\right\rangle _{B}\right) \\
\Phi _{E}^{\pm } &=&\frac{1}{\sqrt{2}}\left( \left| 0\right\rangle
_{A^{\prime }}\left| 0\right\rangle _{B^{\prime }}\pm \left| 1\right\rangle
_{A^{\prime }}\left| 1\right\rangle _{B^{\prime }}\right) ,\Psi _{E}^{\pm }=%
\frac{1}{\sqrt{2}}\left( \left| 0\right\rangle _{A^{\prime }}\left|
1\right\rangle _{B^{\prime }}\pm \left| 1\right\rangle _{A^{\prime }}\left|
0\right\rangle _{B^{\prime }}\right)  \nonumber
\end{eqnarray}

Equation (1) shows how the original entanglement condition existing within
the two separated systems $(k_{A},k_{A^{\prime }})$ and $(k_{B},k_{B^{\prime
}})$ can be swapped to the ''extreme'' modes $k_{A}$ and $k_{B}$ by any
joint Bell type measurement made by Eve on the ''intermediate'' modes $%
(k_{A^{\prime }},k_{B^{\prime }})$. In absence of such a measurement the
overall state $\left| \Phi \right\rangle $ is a superposition while the one
reaching the $(A+B)$ sector is a mixed state.

Suppose that one of the two detectors $D_{j}$ of the Eve sector ''clicks'',
i.e. measures the state $\Psi _{E}^{-}$, say. A sudden state reduction
occurs that projects the overall system onto the corresponding entangled
Bell state: $\left| \Phi \right\rangle \Rightarrow \Psi ^{-}$. The Eve's
apparatus consisting of a 50:50 beam splitter and of a $\varphi $ - phase
shifter is apt to perform this task with a 50 \% efficiency. Indeed it can
be easily found by applying the standard BS theory \cite{5} that the
realization of the 1-photon Bell state $\Psi _{E}^{-}$ (or $\Psi _{E}^{+}$)
over the input modes $k_{A^{\prime }}$, $k_{B^{\prime }}$ determines a click
by $D_{1}$ (or $D_{2}$ ). It is also well known that the states $\Phi
_{E}^{\pm }$ corresponding to a 2 photon excitation of the Eve's sector
cannot be discriminated by any linear device \cite{19}. Note however that
the present experiment is {\it noise free} since a 2-photon excitation of
Eve's sector implies no detections by the $(A+B)$ sector, an event easily
discarded by the electronic apparatus. An additional degree of freedom under
Eve's control, indeed an optional ''delayed choice'', was provided by the
micrometric displacement $\Delta X$ of the mirror $M$, activated by a
piezoelectric transducer. This one induced a corresponding phase shift $%
\Delta \varphi =(2)^{3/2}\pi \lambda ^{-1}\Delta X$ between the modes $%
(k_{A^{\prime }},k_{B^{\prime }})$. Optionally, the same task can be
accomplished by fast Electro Optic (EO) phase modulator, as we shall see.
The 4 detectors adopted in the experiment were equal Si-avalanche
EG\&G-SPCM200 modules with quantum efficiencies: $QE\approx 0,45$.

Suppose that a complete EPR non-locality test is performed by Alice and Bob
by means of\ the two optical homodyne devices shown in Fig. 1, according to
the scheme by Tan et al. \cite{11}. Assume that the eigenvalues of the
clock-LO coherent states are: $\alpha =|\alpha |\exp \theta $, $\alpha
^{\prime }=|\alpha |\exp \theta ^{\prime }$. By a simple extension of a
previous analysis \cite{11} it can be shown that if Eve's detector $D_{1}$
clicks, i.e. $\Psi ^{-}$ is realized, the probability of a coincidence
involving the detectors $D_{A}^{\prime }$ and $D_{B}$ is: 
\begin{equation}
\left\langle \Psi ^{-}\left| I_{A^{\prime }}I_{B}\right| \Psi
^{-}\right\rangle =\frac{1}{4}\left| \alpha \right| ^{2}\left\{ \left|
\alpha \right| ^{2}+\left[ 1+\cos \left( \theta ^{\prime }-\theta +\varphi
\right) \right] \right\}
\end{equation}

Rather than performing the difficult double homodyne experiment, in our case
Alice and Bob carried out an equally significant EPR non-locality test by
mixing the modes $(k_{A},k_{B})$ by a 50:50 BS coupled to the detector pair $%
D_{1}^{\ast }$, $D_{2}^{\ast }$:$\;$Figure 2, {\it inset}. In analogy with
Eve's apparatus, this device may be thought to perform a test on the Bell
states $\Psi ^{\pm }$\ spanning the Hilbert subspace pertaining to the 2-d
manifold $(k_{A},k_{B})$. At the same time it also provides the necessary
synchronizing {\it clock} effect, as said. Consider for instance the
photo-detection by $D_{1}^{\ast }$. Note first that the coincidence
probability of simultaneous clicks by $D_{1}^{\ast }$ and $D_{1}$ is found
by standard theory to be expressed by\cite{5}: 
\[
\left\langle \Psi ^{-}\left| I_{D}I_{D^{\ast }}\right| \Psi
^{-}\right\rangle =\frac{1}{2}\left[ 1+\cos \varphi \right] 
\]
proportional to the expression (3) obtained for the homodyne devices by
setting $\theta =\theta ^{\prime }$ and $|\alpha |^{2}<<1$. Similar results
are found for the other three coincidence combinations involving $%
D_{i}^{\ast }$ and $D_{j}$ $(i,j=1,2)$. Let Alice and Bob carry an
experiment aimed at the measurement of the rate of detection by $D_{1}^{\ast
}$: Figure 2. Since the two systems to be tested $(k_{A},k_{B})$ lack of any
original non-local character, it is natural to expect a total insensitivity
to any change of local parameters acting on remote parts of the apparatus,
as for instance the phase shift $\Delta \varphi $. This is indeed shown by
the experimental data (open circles) given in Figure 2. However, had the
''verification party'', Victor kept the record of the individual outcomes of
both pairs $(D_{1},D_{2})$ and $(D_{1}^{\ast },D_{2}^{\ast })$, at a later
time he could sort into two subsets the already tested samples detected by
Alice and Bob. Figures 2 and 3 show that indeed each subset behaves as if it
consisted of entangled pairs of distant systems. Note that these ones have
never communicated in the past even indirectly via other systems.
Furthermore, as pointed out by Peres, after Alice and Bob have recorded the
results of all their measurement, Eve has still the freedom of deciding
which experiment she will perform \cite{8}. This one may consist of a
standard Bell measurement, or a joint measurement with a $\Delta \varphi $
shift, or a POVM measurement \cite{18} or one of the exotic, interesting
single-photon non-locality tests suggested by L. Hardy \cite{12}. Indeed in
the present experiment, owing to a spatial displacement of the corresponding
detector sets, Eve's action could take place with a time delay $\Delta \tau
\approx 3ns>>\tau _{c}$ respect to the time of the state reduction event
determined by the test performed by Alice and Bob. In other words, since in
our case $\Delta \tau $ was about $3\times 10^{3}$ larger than $\tau _{c}$,
the photon ''{\it coherence time}'', the ''swapping'' process was completed
by the Eve's apparatus long after the complete annihilation of the particle
measured by $(A+B)$.\ In order to offer a even more convincing
demonstration, a sophisticated $\Delta \tau =20ns$ delay apparatus\ has been
realized allowing delayed fast $\Delta \varphi $ changes by a randomly
driven EO Phase Modulator (Inrad 621-040 with: $\Delta \varphi =\pi \equiv
\Delta \varphi _{\lambda /2}$ driven by $400V$ rectangular pulses) triggered
by the $(A+B)$ detection apparatus, i.e. long {\it after} the completion of
the A+B\ test. Note in Figure 2 that shifts $\pm \Delta \varphi _{\lambda
/2} $ correspond to the detection interchanges: $\Psi ^{-}$ $%
\rightleftarrows \Psi ^{+}$. This makes the original Peres's argument,
conceived for standard Bell-inequality tests of $(2\otimes 2)$-d Hilbert
photon $\pi $-states, fully consistent with the present experiment \cite{1,8}%
.

How could then Eve's delayed choice determine data already irrevocably
recorded ? According to Peres, it is meaningless to assert that two quantum
systems are entangled without specifying their state, or to assert that a
system is in a pure state without specifying that state or to attribute an
objective meaning to the quantum state of a single system. If these
prescriptions are forgotten one may encounter paradoxes as the one seen
here: a past event may sometimes appear to be determined by future actions 
\cite{8}. For better clarification it is perhaps worth reminding here that:\
''any phenomenon is not a phenomenon until is a measured phenomenon'' (J.
Wheeler) asserting the inanity of any intellectual speculation involving
mental modelling of the inner evolution of\ a quantum superposition process.
Furthermore, as pointed out by Richard Jozsa \cite{15}, any apparent
retrodictive process, e.g. associated with the quantum evolution in presence
of the EPR\ nonlocality in a teleportation process, cannot finally lead to
paradoxes or contradictions of causality because of the inherent
inaccessibility of the quantum information.

In conclusion, we have illustrated experimentally an enlightening aspect of
quantum EPR non-locality. We have accomplished that by implementing a new
method of photon quantum entanglement that is expected to play a significant
role in the field of modern quantum information as well in future studies on
EPR quantum non-locality. For instance the application of our new methods of
\ high ''fidelity''quantum teleportation and entanglement swapping to modern
''quantum repeaters'' will certainly improve in the near future the
technology of quantum communication at large distances \cite{14,20}.

We are indebted with the FET European Network on Quantum Information and
Communication (Contract IST-2000-29681-ATESIT) and with M.U.R.S.T. for
funding.

\centerline{\bf Figure Captions}

\vskip 8mm

\parindent=0pt

\parskip=3mm

FIG. 1. Layout of an experimental demonstration of the delayed-choice
entanglement-swapping process. In the actual experiment the 2-homodyne
apparatus was replaced by the 2-detector set shown in Figure 2, inset.

FIG. 2. Experimental results of the measurement of the count rate by
detector $D_{1}^{\ast }$ as function of delayed settings of the phase $%
\varphi $ determined by micrometric displacements $X$ of the mirror $M$
(open circles). A verification party, Victor can sort at a later time the
recorded pattern in two subsets showing two sinusoidal fringe patterns with
opposite phases corresponding to the Bell states $\Psi ^{\pm }$\ , for $%
\varphi =0$. The visibility of the fringe patterns is $V=91\pm \ 2\%$. The
inset shows the 2-detector apparatus that has been adopted to perform
experimentally the EPR non-locality test and that for that purpose replaces,
in a fully equivalent fashion the double homodyne apparatus shown in Figure
1.

FIG. 3. Histograms showing the measured detection count rates by $%
D_{1}^{\ast }$ and $D_{2}^{\ast }$ and the accuracy affecting the
experimental determination of the Bell states $\Psi ^{\pm }$\ obtained by
the delayed coincidence rates involving all detector pairs: $D_{i}^{\ast }$- 
$D_{j}(i,j=1,2)$, for $\varphi =0$.

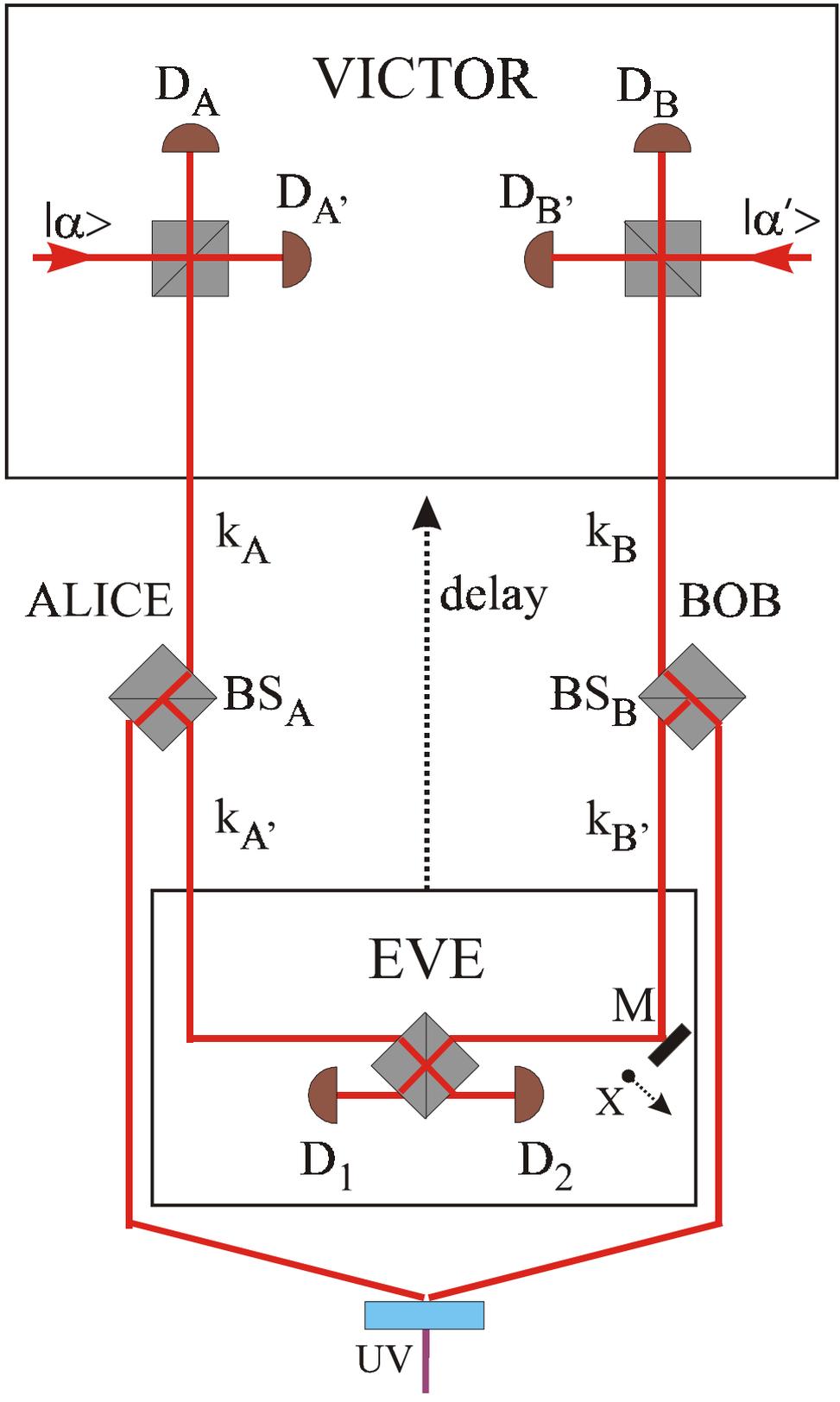
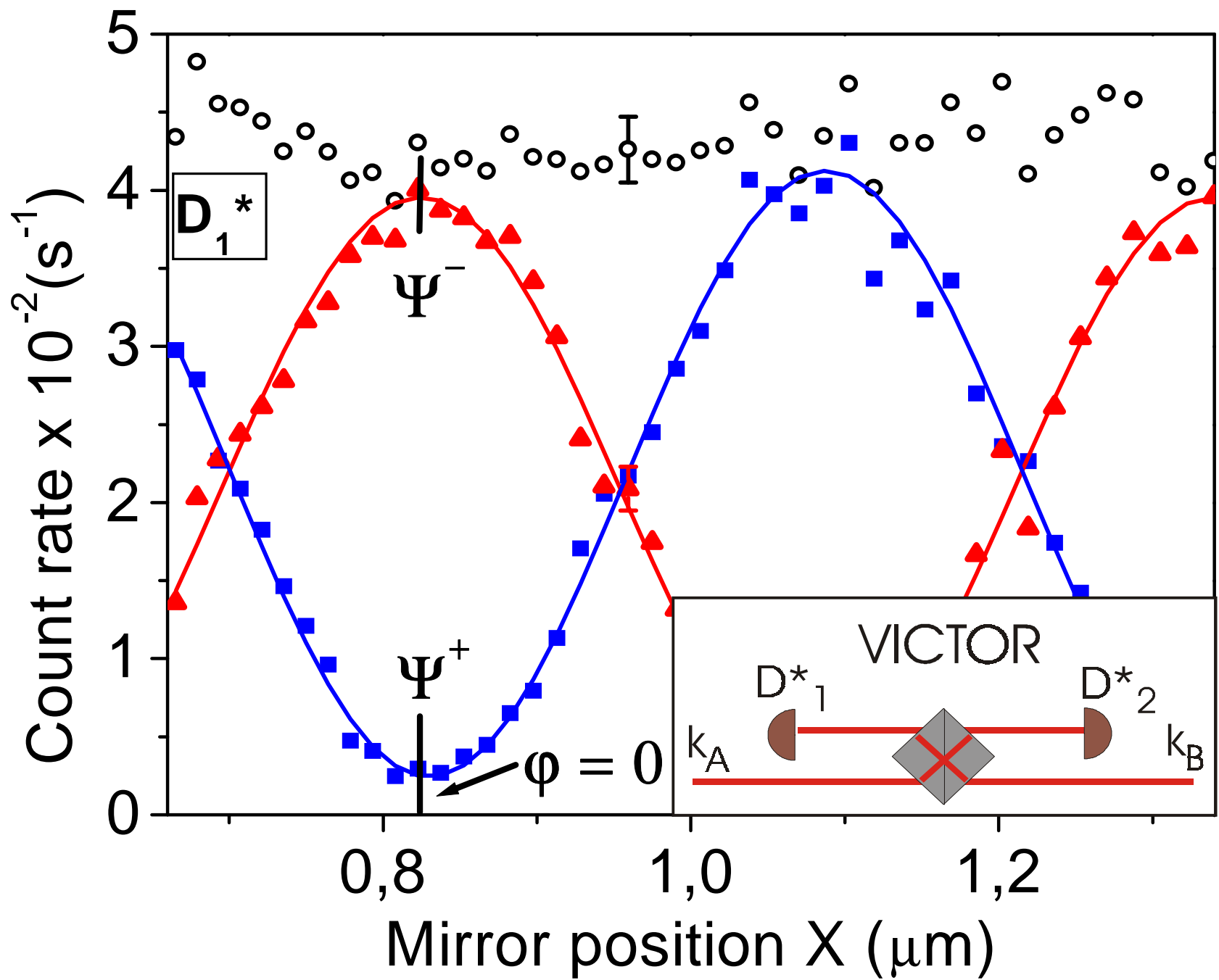
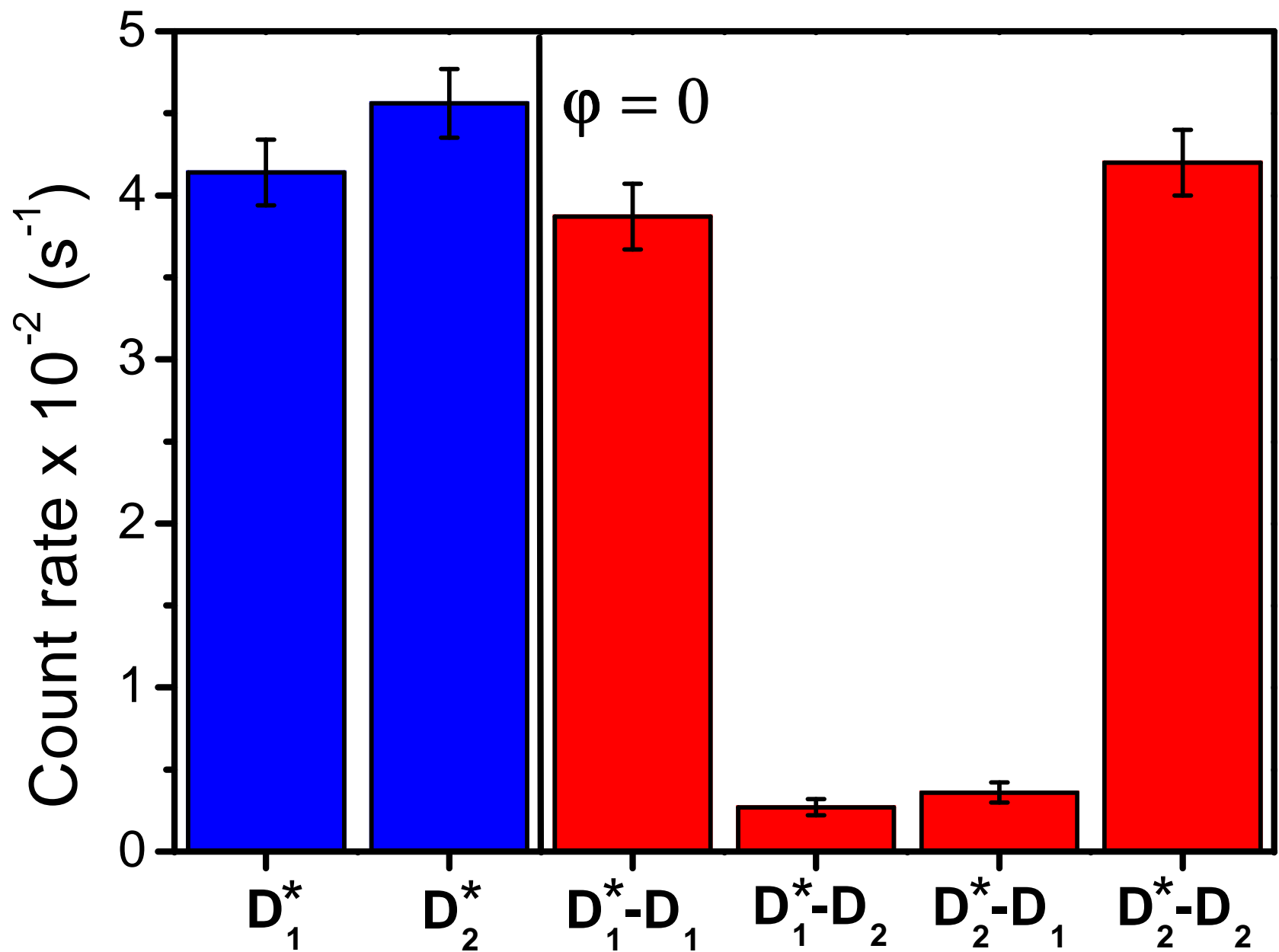
\end{document}